\documentclass[final]{svjour3}
\usepackage[utf8]{inputenc}
\usepackage{color}
\usepackage{graphicx}
\usepackage{rotating}
\usepackage{amssymb}
\usepackage{mathptmx}
\usepackage[numbers]{natbib}
\usepackage{wrapfig}
\usepackage{graphicx}
\usepackage[dvipdfm,colorlinks=true,allcolors=blue]{hyperref}
\usepackage{amsmath}
\makeatletter 
\journalname{Arxiv}
\begin{document}
\author{Christopher B. McKitterick \and Heli Vora \and Xu Du \and Boris S. Karasik \and Daniel E. Prober}
\institute{Christopher B. McKitterick \and Daniel E. Prober \at Departments of Physics and Applied Physics, Yale University, New Haven, Connecticut 06520, USA\\
\email{chris.mckitterick@yale.edu}\\
\email{daniel.prober@yale.edu}
\and Heli Vora \and Xu Du \at Department of Physics, Stony Brook University, Stony Brook, New York 11790, USA
\and Boris S. Karasik \at Jet Propulsion Laboratory, California Institute of Technology, Pasadena, California 91109, USA}
\title{Graphene microbolometers with superconducting contacts for terahertz photon detection}
\maketitle

\begin{abstract}
We report on noise and thermal conductance measurements taken in order to determine an upper bound on the performance of graphene as a terahertz photon detector. The main mechanism for sensitive terahertz detection in graphene is bolometric heating of the electron system. To study the properties of a device using this mechanism to detect terahertz photons, we perform Johnson noise thermometry measurements on graphene samples. These measurements probe the electron-phonon behavior of graphene on silicon dioxide at low temperatures. Because the electron-phonon coupling is weak in graphene, superconducting contacts with large gap are used to confine the hot electrons and prevent their out-diffusion.  We use niobium nitride leads with a $T_\mathrm{c}\approx 10$~K to contact the graphene. We find these leads make good ohmic contact with very low contact resistance. Our measurements find an electron-phonon thermal conductance that depends quadratically on temperature above 4~K and is compatible with single terahertz photon detection.
\end{abstract}

\section{Introduction}
Graphene has been proposed for use in many applications, ranging from solar cells \cite{Wang2008} to photon detectors \cite{Vora2011,Yan2011,Betz2012a,Fong2012a}. The low heat capacity \cite{Fong2012a} of graphene makes it an appealing candidate for detecting low energy photons. In particular, a detector of far-IR terahertz (THz) photons is of great interest for both laboratory experiments \cite{Schmutt} and astronomical studies in space-based observatories \cite{Karasik}. Transition edge sensors (TESs) using the superconducting transition of titanium (Ti) have been proposed as a detector for photons in the far-IR \cite{Karasik2005}. These TES detectors have achieved single-photon sensitivity in the near- and mid-IR \cite{Lita2008,Karasik}. Quantum capacitance\cite{Stone2012} and kinetic inductance \cite{Day2003,mkid} detectors are also emerging as candidates for detecting single THz photons. 

Unlike a transition edge sensor biased on its superconducting transition, the resistance of graphene is not significantly temperature dependent for the conditions where it is most promising for photon detection. An alternate readout mechanism must be employed. We have previously modeled the use of Johnson noise to measure the absorption of THz photons in graphene \cite{McKitterick2013}. This method of reading out the electron temperature introduces an additional source of noise \cite{Dicke1946}, Eq.~\ref{eq:dicke} below, that must be considered when evaluating detector performance. To determine the sensitivity of the graphene detector, the change in temperature due to an incident THz photon, $\Delta T$, must be compared to the uncertainty in the temperature measurement, $\delta T$. 

There are two dominant contributions to the temperature uncertainty: intrinsic temperature fluctuations due to the exchange of energy (e.g., phonons, photons, hot electrons) between the graphene and its environment($\delta T_\mathrm{intr}$) \cite{Mather1984} and apparent temperature fluctuations due to the inaccuracy of the Johnson noise-readout ($\delta T_\mathrm{readout}$). These sources of noise add in quadrature and are presented below:\\
\begin{minipage}{.5\linewidth}
\begin{equation}
\label{eq:intr}\delta T_\mathrm{intr}\approx\sqrt{\frac{k_\mathrm{B}(T^2+T_0^2)}{2C}}
\end{equation}
\end{minipage}%
\begin{minipage}{.5\linewidth}
\begin{equation}
\label{eq:dicke}\delta T_\mathrm{readout}=\frac{T_\mathrm{a}+T}{\sqrt{B\tau}}.
\end{equation}
\end{minipage}
Here, $T$ denotes the electron temperature, $T_0$ is the substrate temperature, $C$ is the heat capacity, $T_\mathrm{a}$ is the noise temperature of the amplifier, $B$ is the measurement bandwidth, and $\tau$ is the time interval over which the temperature is being averaged. With a smaller thermal conductance, the thermal relaxation time of the system is increased, allowing the noise due to the readout to be reduced by averaging for a longer time. For graphene to be a practical detector, it is critical to achieve very low values of thermal conductance.

There are additional constraints on the device. Graphene samples need to be fabricated with a very low heat capacity and used at a low bath temperature ($T_\mathrm{0}\sim 100$~mK) in order to realize the necessary sensitivity to detect single THz photons. Operating with this low heat capacity, however, means that an incident THz photon will cause a temperature rise that is significantly larger than the equilibrium temperature of the bath ($T\gg T_\mathrm{0}$) so the standard equilibrium formula for Eq.~\ref{eq:intr} that takes $T=T_0$ is no longer valid. In this case, where $T\gg T_0$, the temperature used to calculate the noise of the system for a particular averaging interval is taken to be the average electron temperature over that interval, $T=T_\mathrm{avg}$.

\begin{wrapfigure}{r}{0.5\textwidth}
 \centering
 \includegraphics[width=.48\columnwidth]{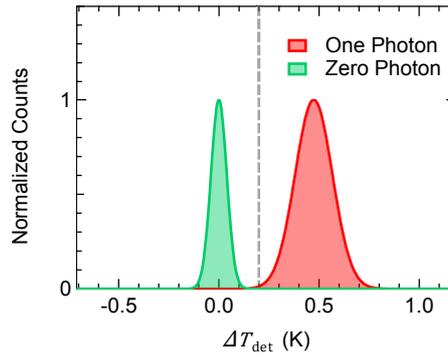}
 \caption{(Color online) Normalized histograms for photon counts using a previously reported thermal conductance \cite{Betz2012} scaled to 100~mK and a heat capacity of $2\times 10^{-22}~$J/K at 100~mK. The amplifier used for these calculations is described in McKitterick et al. \cite{McKitterick2013}.}
\label{fig:hista}
\end{wrapfigure}

By using previously measured values for thermal conductance \cite{Betz2012} and heat capacity \cite{Fong2012a}, and extrapolating to lower temperatures, we performed calculations to determine the ideal operating conditions for a graphene detector. We find that best detection of single photons occurs for parameters such that $T_\mathrm{avg}\gg T_0 $. An approximate upper bound on heat capacity of $2\times 10^{-22}$~J/K at 100~mK is necessary to achieve good performance in detecting single THz photons \cite{McKitterick2013}. This value of heat capacity corresponds to a 4~$\mu\mathrm{m}^2$ flake of graphene with a carrier density of $n=10^{12}~\mathrm{cm}^{-2}$. In Fig.~\ref{fig:hista}, we present histograms of theoretical photon counts of single 1~THz photons for a hypothetical graphene detector with these characteristics; this shows the ensemble behavior 

In Fig.~\ref{fig:hista}, the histogram on the right represents the relative probability of reading out a certain change in temperature, $\Delta T_\mathrm{det}$, when a 1~THz photon arrives at the detector. Similarly, the histogram at the left represents the relative probability of observing an apparent change in temperature when no photons are absorbed in the averaging interval. If this detector had no noise, these histograms would resemble delta functions and be centered either at the average temperature change due to the arrival of a photon, or at zero detected temperature change. The broadening of the histograms is due to the contributions of the noise outlined in Eqs.~\ref{eq:intr} and \ref{eq:dicke}.

The detection peak for single-photon events is well-separated from the histogram of counts when no photons arrive at the detector. Using the calculated thermal relaxation time of $\tau=0.5~\mu$s and setting a threshold on the minimum change in electron temperature detected, $\Delta T_\mathrm{det}$, of 200~mK, one can limit the dark (zero-photon) counts to the background count rate, taken to be 1000 photons per second \cite{Karasik2011a}, without losing many single-photon events. This is an effective photon counter. With a cold, tunable frequency pre-filter, this graphene photon-counting detector could provide THz spectroscopy at the single-photon level \cite{Karasik2005}.

In order to determine whether the parameters assumed are valid, it is necessary to measure how the thermal conductance behaves at low bath temperatures. We assume there are three cooling pathways for the hot electrons: coupling to phonon modes ($G_\mathrm{el-ph}$), emission of microwave photons ($G_\mathrm{photon}$) \cite{Schmidt2004}, and out-diffusion of hot electrons ($G_\mathrm{diff}$), yielding a total thermal conductance of $G_\mathrm{tot}=G_\mathrm{el-ph}+G_\mathrm{photon}+G_\mathrm{diff}$. 

To achieve the low value of thermal conductance that is necessary for THz photon detection, the cooling pathway due to hot electrons diffusing out the leads needs to be suppressed and the electron-phonon coupling must be very weak at low temperatures. Superconducting contacts have been proposed to virtually eliminate $G_\mathrm{diff}$ at low temperatures, as demonstrated for lead (Pb) contacts \cite{Borzenets2011}. However, Pb is unsuitable for the applications proposed herein due to its formation of a thick oxide barrier. Effective confinement of electrons in graphene by other large-gap superconductors has not previously been reported.

Theoretical predictions for the electron-phonon coupling suggest a power law form for the electron-phonon cooling, $G_\mathrm{el-ph}\propto T^p$, with $p$ equal to 2, 3, or 4 \cite{Song2011,Chen2012,Tse2009}. Previous measurements of electron-phonon thermal conductance at cryogenic temperatures have yielded inconsistent results. Two groups \cite{Fong2012a,Betz2012} have found $G_\mathrm{el-ph}\propto T^3,$ above 2~K but their extracted values for electron-phonon coupling strength differ by over two orders of magnitude. Other studies have observed a $T^2$ dependence for electron-phonon thermal conductance from 10 to 300 ~K \cite{graham2012photocurrent,Betz2012a} and 100~mK to 700~mK \cite{Borzenets2012}. 

In this article, we present preliminary measurements on graphene fabricated on silicon dioxide (SiO$_2$) that aim to provide a better understanding of $G_\mathrm{el-ph}$ and establish the capability of superconducting contacts on graphene to block $G_\mathrm{diff}$.

\section{Methods}
\begin{figure}[]\sidecaption
 \centering
 \includegraphics[width=.6\columnwidth]{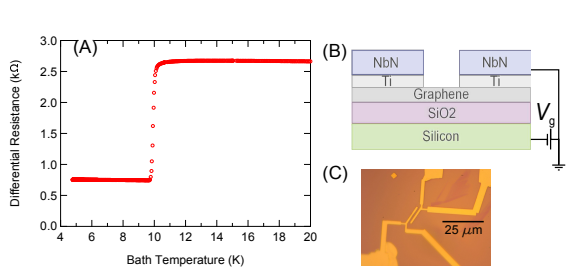}
 \caption{(Color online) (A) dV/dI as a function of temperature measured with a $1~\mu$A AC bias. Above 10~K, the resistance is primarily due to the NbN leads. (B) Schematic of graphene device (not to scale). A gate voltage is used to tune the carrier density and Ti is used as an adhesion layer for the NbN. (C) Optical micrograph of NbN lead structure. The spacing between the leads is approximately 1$~\mu$m.}
\label{fig:gsamp}
\end{figure}
The graphene samples used in this study are fabricated at Stony Brook University, where graphene is exfoliated onto a silicon substrate (15-20~$\Omega\cdot$cm at 300~K) with a 500~nm thick layer of SiO$_2$. Niobium nitride (NbN) leads are deposited on the graphene flake with a Ti adhesion layer. A thin layer of Pd is deposited between the Ti and NbN to reduce interaction between the two layers during deposition. The sample presented here has a NbN transition temperature of approximately 10~K with graphene dimensions $10\times 4~\mathrm{\mu m}$ ($10\times 1~\mathrm{\mu m}$ between leads). This transition is due to the superconducting resistance change of the NbN leads. 

The samples are mounted in a vacuum can that is cooled to approximately 2~K in a pumped liquid helium cryostat. Measurements of thermal conductance are performed by heating up the electrons with a DC current and using the microwave readout shown in Fig.~\ref{fig:tcon}A to measure the emitted Johnson noise.
The Johnson noise is rectified using a diode to present a DC voltage to a digital multimeter. To convert this voltage to the electron temperature, we set $V_\mathrm{DC}=0$ so $T=T_0$ and increase the temperature of the graphene sample using a resistive heater and record the resulting linear change in voltage read out by our diode. 

Using this calibration, we find the electron temperature as a function of the DC heating power due to an applied current. We then use the electron temperature to determine the thermal conductance, $G=\frac{dP}{dT}.$
We present this in Fig.~\ref{fig:tcon}B along with the calculated value for $G_\mathrm{diff}$ using the sample resistance to determine $G_\mathrm{diff}$. 
\section{Discussion}
For large values of electron temperature, adding a power law term to $G_\mathrm{diff}$ describes the thermal conductance fairly well and yields a power of 2 for the fit parameter $p$. This is consistent with recent theory \cite{Song2011} predicting that disorder-assisted scattering, described as supercollisions, can dominate the electron-phonon cooling, which results in a $G_\mathrm{el-ph}\propto T^2$. The disorder-assisted scattering theory predicts a crossover temperature, $T_*$, below which $G_\mathrm{el-ph}\propto T^3$. In that article, the authors expect $T_*\gtrsim 10$~K for typical carrier densities. For our data, the large contribution of $G_\mathrm{diff}$ below approximately 8~K overrides the contribution of electron-phonon cooling at these temperatures, preventing the extraction of $T_*$.
\begin{figure}[]
 \centering
 \includegraphics[width=.8\linewidth]{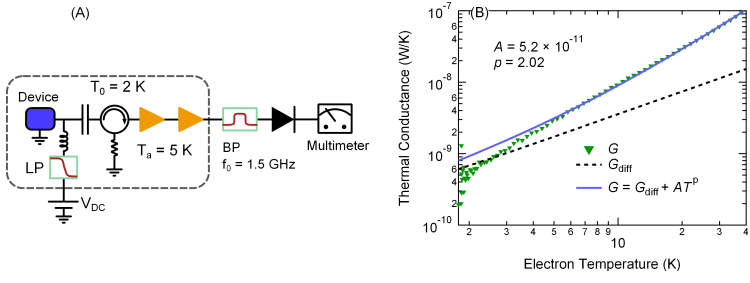}
 \caption{(Color online) (A) Schematic of experimental setup for Johnson-noise measurements. LP and BP stand for low-pass and band-pass respectively. (B) Calculated thermal conductance from measured electron temperature. $G_\mathrm{diff}=\frac{12\mathcal{L}T}{R}$ is calculated from the measured electron temperature and device resistance. $\mathcal{L}=2.44\times 10^{-8}~\mathrm{W\Omega K^{-2}}$ is the Lorentz number and $R$ is the electrical resistance \cite{Prober1993}. Here $G_\mathrm{diff}$ is computed assuming non-superconducting contacts.}
\label{fig:tcon}
\end{figure}

At temperatures lower than 5~K, this fit ceases to accurately describe the measured thermal conductance. We believe that the significant drop of the measured $G$ below the fit line is the result of the superconducting contacts blocking the hot electrons from diffusing out of the graphene. 
\begin{figure}[]
 \centering
 \includegraphics[width=.8\linewidth]{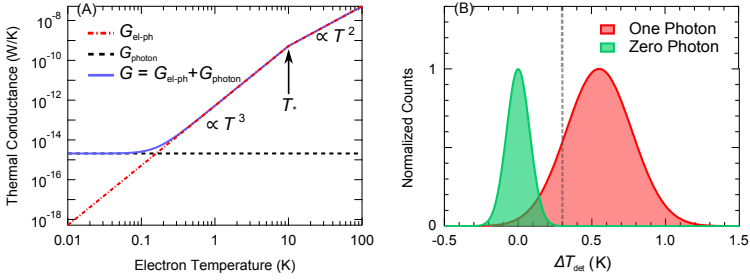}
 \caption{(Color online) (A) Estimated electron-phonon contribution to thermal conductance, assuming $T_*=10$~K and an effective area of $4~\mathrm{\mu m}$. $G_\mathrm{rad}$ is calculated using a coupled bandwidth of 150~MHz. (B) Normalized histograms for photon counts using the thermal conductance from (A) and a heat capacity of $2\times 10^{-22}~$J/K at 100~mK. The dashed line indicates a threshold $\Delta T_\mathrm{det}$ which would have few dark counts, but would count approximately $85\%$ of the single-photon events.}
\label{fig:hist-new}
\end{figure}

For the purposes of THz detection, we now consider a similar graphene flake to the one measured in Figs.~\ref{fig:tcon} and \ref{fig:hist-new}, but with an area ten times smaller, corresponding to the optimal heat capacity found in previous calculations \cite{McKitterick2013}. At 100~mK,  a device with these dimensions may permit a Josephson current \cite{Borzenets2012}. That would not be desirable, due to the excess noise at finite DC voltages needed to read out the noise. In order to couple to the device, it may be necessary to use superconductor-insulator-graphene (SIG) contacts with a thin tunnel barrier to suppress the Josephson current. For Johnson-noise readout at microwave frequencies, the finite impedance of the tunnel barriers would be short circuited by the tunnel junction capacitance. The device would be impedance matched to the readout circuit.

To evaluate the performance at 100~mK, we made several assumptions about the behavior of the thermal conductance below 1~K. First, we assume that $G_\mathrm{diff}\rightarrow 0$ for $T<1$~K so that $G=G_\mathrm{el-ph}+G_\mathrm{photon}$, where $G_\mathrm{photon}=k_\mathrm{B}B$ and $B$ is the coupled bandwidth to an impedance-matched load. We assume that $G_\mathrm{el-ph}\propto T^3$ below 10~K; this is shown in Fig.~\ref{fig:hist-new}. From these assumptions, the cooling power due to electron-phonon coupling is given by $P_\mathrm{el-ph}=\frac{A}{4T_*}(T^4-T_0^4)$. We can then use the heat flow equation to calculate the temperature of the electrons as a function of time after absorbing a THz photon at 100~mK, as outlined \cite{McKitterick2013}. From the calculation of electron temperature as a function of time, we determine the rms apparent temperature fluctuations, $\delta T$. The expected histograms of 1~THz photon counts are shown in Fig.~\ref{fig:hist-new}.

This detector would have a sufficient resolving power to count single 1~THz photons. By setting a threshold $\Delta T_\mathrm{det} = 300$~mK, this detector would be limit the dark counts to the background count rate (taken to be 1000 photons per second \cite{Karasik2011a}) while still recording approximately $85\%$ of single-photon events. With a well matched antenna, this detector would achieve a single-photon detection efficiency for 1~THz photons much greater than any device reported so far \cite{Komiyama}. 

There are still open questions about the cooling of hot electrons in graphene. Previous measurements above 1~K are not in agreement. Also, it is not clear what area of the graphene contributes to the electron-phonon thermal conductance. At high temperatures, $G_\mathrm{el-ph}\gg G_\mathrm{diff}$, and the effective area should be much smaller than the total area. At low temperatures, a larger area may contribute. Finally, the effect of the substrate on electron-phonon cooling is still not well understood. These questions define the goals for future studies.

\section{Conclusions}
The measurements we present here are consistent with graphene being a good single-photon THz detector. To really evaluate the effectiveness of this type of detector, measurements need to be made down to the base temperature of 100~mK. Such measurements would serve both to provide a better understanding of the electron-phonon cooling and would demonstrate the capability of superconducting contacts to effectively confine hot electrons. Both of these goals are critical to the design of a graphene THz photon detector.

{\footnotesize
\bibliographystyle{Science}
}

\begin{acknowledgements}
The work at Yale was supported by NSF Grant DMR-0907082, an IBM Faculty Grant, and by Yale University. The work at Jet Propulson Laboratory was carried out under a contract with the National Aeronautics and Space Administration. X. Du and H. Vora acknowledge support from AFOSR-YIP Award No. FA9550-10-1-0090. 
\end{acknowledgements}
\end{document}